\documentclass[pre,twocolumn,showpacs,preprintnumbers,amsmath,amssymb,floatfix]{revtex4}

\usepackage{graphicx}

\begin{document}

\title{Survival probability and order statistics of diffusion on disordered media
}
\author{L. Acedo}
\email[E-mail: ]{acedo@unex.es}

\author{S. B. Yuste}
\email[E-mail: ]{santos@unex.es}
\affiliation{%
Departamento de F\'{\i}sica, Universidad  de  Extremadura,
E-06071 Badajoz, Spain
}%

\date{\today}
\begin{abstract}
We investigate the first passage time
$t_{j,N}$ to a given chemical or Euclidean distance of the first $j$ of a set of $N\gg
1$ independent random walkers all  initially placed on a site of a disordered medium.
To solve this order-statistics problem we assume that, for
short times, the survival probability (the probability that a single random walker is
not absorbed by a hyperspherical surface during some time interval) decays  for
disordered media in the same way as for Euclidean and some class of
deterministic fractal lattices. This conjecture is checked by simulation on the incipient percolation aggregate embedded in two dimensions. Arbitrary moments of
$t_{j,N}$ are expressed in terms of an asymptotic series in powers of $1/\ln N$ which is
formally identical to those found for Euclidean and  (some class of) deterministic fractal
lattices. The agreement  of the asymptotic expressions with simulation results for the
two-dimensional percolation aggregate is good when the boundary is defined in terms of
the chemical distance. The agreement worsens slightly  when the Euclidean distance is used.

 \end{abstract}

 \maketitle

\section{INTRODUCTION}
\label{sec:Introduction}

Diffusion in disordered media has been an area of intensive research during the last two decades \cite{ReviewHB,BookBH}.
Transport in non-crystalline, disordered materials cannot be explained by the
classical theories of diffusion since anomalous behaviors  (relative to what happens
in the Euclidean domain) are the rule here.
In particular, the mean-square displacement of a random walker $\langle r^2 (t) \rangle$ is no longer proportional to the time $t$ as occurs in uniform Euclidean systems of any dimension (Fick's Law), but the more general anomalous diffusion law $\langle r^2 \rangle \sim 2 D t^{2/d_w}$ holds for large times, $r$ being the Euclidean distance from the position of the random walker at $t=0$ ,  $d_w$  the anomalous diffusion exponent, and $D$  the diffusion constant.
As the geometrical structure of real disordered media is very complex and varied, it is usually
modeled by stochastic fractal lattices. 
Of these, the incipient percolation aggregate embedded in either two or
three dimensions is the most widely used \cite{ReviewHB,BookBH,BookSA,WeissEdHB,Essam}.

Statistical problems related to a single random walker have
traditionally been the subject of more intensive research than
those corresponding to $N > 1$ interacting or independent random
walkers \cite{Hughes,Weiss}. 
Of course, problems in which the walkers interact can not be analyzed  in terms of the single-walker theory. 
However, there also exist other multiparticle problems that can not be analyzed in terms of the single-walker theory even though the walkers are independent. 
These problems have begun to be the target of in-depth studies  during the last decades
\cite{NAT,PRAHL,JCPSA,SNTEuc,SNTFrac,WSL,PRLYuste,JPA2000,YAL,JPAKR,PREDK}.
Particular attention has been paid to (i) the evaluation of the average number of distinct sites visited up to time
$t$ by $N$ independent random walkers all starting from the same
origin in both Euclidean and fractal lattices \cite{NAT,PRAHL,JCPSA,SNTEuc,SNTFrac}, and (ii) the description of the order statistic of the diffusion processes. This is the subject we address in this paper and can be stated as follows. A set of $N$ independent random walkers all initially placed at a given site (the origin) of a medium start to diffuse at time $t=0$.
Eventually, at time $t_{1,N}(z)$ a random walker of this set reaches for the first time a site separated from the
origin  by the distance $z$.
Next, a second random walker reaches at time $t_{2,N}$  a site at the same distance $z$ from the origin,  and so on \cite{SingMol}.
Equivalently, we can understand  $t_{j,N}(z)$ as the time taken (escape time or lifetime) by the  $j$th particle out of a set of  $N\gg 1$ to  escape  from a ``spherical'' region of radius $z$ centered  at the starting site of diffusion.   
Our goal in this paper is to  calculate the escape-time moments $\langle t^p_{j,N}(z)\rangle$
when $N\gg 1$ random walkers diffuse  in a {\em disordered}  medium. 

This order-statistic  problem was solved  for the one-dimensional lattice in Ref.\ \cite{WSL,PRLYuste} and   for some class of {\em deterministic}   fractal substrates in  Ref.\ \cite{PRLYuste}.
For  $d$-dimensional Euclidean lattices,   the form of the first moment of  $t_{1,N}$  was guessed  in Ref. \cite{JPA2000} and checked using simulations results for $d=2$ and $d=3$.
The full solution of this problem for Euclidean media and for arbitrary  $p$ and $j$ has been obtained recently \cite{YAL}.
For all these media,  the $p$th moment of the time $t_{j,N}$ spend by the $j$th random walker in reaching the
Euclidean distance $r$  was given in terms of an asymptotic series for large $N$ of the
form  
\begin{equation}
\langle t^p_{j,N}\rangle \sim
 \left(\frac{r}{\sqrt{2D}}\right)^{p d_w}(\ln N)^{p(1-d_w)}
      \sum_{n=0}^\infty  \sum_{m=0}^n \tau_{nm} \frac{(\ln\ln N)^m}{(\ln N)^{n}} ,
\label{eq:soluformal}
\end{equation}
where the coefficients $\tau_{nm}$ depend on $p$, $j$ and  the substrate. The question we want to answer is whether this is also true for disordered media.

The prior knowledge of the {\em short-time} asymptotic expression of the survival probability 
$\Gamma(z,t)$, which is defined as the probability that a random walker who starts at an origin site
has not arrived at a spherical boundary of radius  $z$ in the time interval $(0,t)$, was a key that  allowed  to find rigorously the full asymptotic approximation   \eqref{eq:soluformal} for
$\langle t^p_{j,N}\rangle$  for Euclidean and some deterministic fractal media \cite{surviEucDetFrac} in  Refs.\ \cite{WSL,PRLYuste,YAL}.
(Here, and henceforth,  we  use the term ``short time'' or ``short-time regime''  to mean  that  $z/\langle z^2(t) \rangle^{1/2}\gg 1$,
$\langle z^2(t) \rangle$ being the  mean square distance $z$ travelled by a single random walker  by  time $t$.)
However, the use  of the asymptotic procedures of Refs.\ \cite{WSL,PRLYuste,YAL}
for estimating $\langle t^p_{j,N}\rangle$ for {\em disordered} media
is impeded by the fact that  the value of the short-time survival probability is unknown for these substrates.
In spite of this,  in Ref.\ \cite{PRLYuste} it was conjectured that Eq.\ (\ref{eq:soluformal})  is
valid for disordered media too \cite{conjecture}.
However,   Dr\"ager and Klafter  \cite{PREDK}, using a scaling approach, have shown recently that
\begin{equation}
\langle t_{1,N}\rangle =  \mathit{O}\left[ \ell^{d^\ell_w}(\ln N)^{1-d^\ell_w}\right]
\label{eq:soluDK}
\end{equation}
for {\em ordered and disordered} structures, which differs from \eqref{eq:soluformal} because, in general, $d^\ell_w\neq d_w$.
Here $d^\ell_w=d_w/d_{\text{min}}$  is the chemical diffusion exponent defined by the relation
 $\langle \ell  \rangle \sim \text{const}\times t^{1/d_w^\ell}$,  $\ell(r)$  being  the chemical or topological distance defined as the
 length of the shortest path connecting two sites on a substrate that are separated by the Euclidean distance $r$, and
 $d_{\text{min}}$ is the fractal dimension of the shortest path on the fractal:
  $\langle \ell (r) \rangle \sim  \text{const} \times r^ d_{\text{min}}$ \cite{ReviewHB,BookBH}.
  Notice that $d_{\text{min}}=1$ for Euclidean lattices and for the deterministic fractals discussed in Ref. \cite{PRLYuste}, so
  that, for these cases, Eqs. (\ref{eq:soluformal}) and (\ref{eq:soluDK}) agree.
Unfortunately, the scaling approach of Dr\"ager and Klafter  is not precise enough to lead to
a fully correct asymptotic expression of the form of Eq.\   (\ref{eq:soluformal}). 
This can be seen, for example, because the asymptotic series for $\langle t_{1,N} \rangle $ 
given by these authors in Ref.\  \cite{PREDK}  does not agree with the rigorous asymptotic series reported in Refs.\ \cite{WSL,PRLYuste} for the one-dimensional lattice.

In this paper we  deal with the order statistic of the diffusion process in disordered media following the procedure
previously carried out in Euclidean media \cite{WSL,PRLYuste,YAL}
and  in (some class of) deterministic fractal lattices  \cite{PRLYuste,YAL}.
Therefore, our first step must be to propose a short-time asymptotic expression for the
survival probability.  The functional form we take coincides with that obtained for Euclidean and 
renormalizable fractal lattices.
Next, we apply the asymptotic methods already used for the Euclidean and deterministic fractal cases
 in order to obtain asymptotic expansions for the order-statistics moments $\langle t^m_{j,N}\rangle$.

The paper is organized as follows.
 In Sec.\ \ref{sec:SurvivalProbability}, a short-time asymptotic expression for the survival probability for
 disordered media is proposed.  We check this conjecture in the  two-dimensional incipient percolation aggregate and,
 on the way, estimate for this case the unknown parameters appearing in the proposed relationship.
 In Sec.\ \ref{sec:FormulasAsintoticas} we give the asymptotic expressions of the moments of
  $ t_{j,N} $ for stochastic fractals assuming that  the survival probability for the short-time regime
  is given by the expression  proposed in Sec.\  \ref{sec:SurvivalProbability}.
 These theoretical results are compared with simulation
data for the two-dimensional percolation aggregate.
The paper ends with some conclusions and remarks.

\section{SHORT-TIME SURVIVAL PROBABILITY}
\label{sec:SurvivalProbability}

 From now on, we will denote by $z$ either the chemical distance  $\ell$ or the Euclidean distance  $r$.
As discussed in Sec.\ \ref{sec:Introduction}
we have to know the survival probability $\Gamma(z,t)$, or
 equivalently the (boundary) mortality function $h(z,t)=1-\Gamma(z,t)$, for  short times [i.e., for  $z/\langle z^2(t) \rangle^{1/2}\gg 1$] in order to calculate the escape-time moments  $\langle t_{j,N}^m \rangle$.
This  mortality function is the probability that a given random walker starting at the site
$z=0$ will have  been trapped by an absorbing boundary of radius $z$ by time  $t$. 
In this paper we will consider media for which this mortality function (averaged over all realizations
of the lattice if the lattice is stochastic) grows for   $\xi \equiv z/t^{1/d_w^z} \gg 1$ as
a stretched exponential with power-law corrective terms:
\begin{equation}
\label{hzxi}
h(z,t)=h(\xi) \sim A  \xi^{-\mu v} \exp \left[
-c \xi^{v} \right] \left\{1+h_1 \xi^{-v}
+\ldots \right\}
\end{equation}
where  $A$, $\mu$,  $v$,  $h_1$ are characteristic parameters of the lattice.
The anomalous diffusion coefficient
$d_w^z$ is replaced  by $d_w$ if the Euclidean distance is used ($z=r$) and by $d_w^\ell$ if $z=\ell$.
There are good reasons to propose the functional form of Eq.\ (\ref{hzxi}).
To start with, this relation holds for  Euclidean lattices.
Then, the mortality function was obtained in Ref.\ \cite{PRLYuste,JPAYuste} for some class
 of deterministic fractals by
using a renormalization procedure which involved only boundaries containing
the nearest neighbors of the origin after successive decimations,
 and the result is in agreement with  Eq.\ (\ref{hzxi}) too.
 In these media $z \equiv r$,
$\mu=1/2$  and $v=d_w/(d_w-1)$.
(Although the  results of renormalization cannot be directly applied to any origin and any
arbitrary boundary,  there are again good reasons to suspect that the
functional form is the same \cite{Propag}.)
Also, a closely related quantity to $h(z,t)$, the (site)  mortality function  $h_t({\bf r})$ defined as the probability that a specific
site ${\bf r}$ has  been visited by a single random walker  by time t,  follows Eq.\ (\ref{hzxi})  for  the
two-dimensional percolation aggregate \cite{SNTFrac}.
Finally, one expects that the mortality function and the propagator $P(z,t)$ 
 decay for $\xi \gg 1$ in the same way \cite{JPAYuste,Propag}, and it is known
that $P(z,t)$ decays as $\exp(-c \xi ^v)$  for stochastic fractals, where 
$v=d_w^z/(d_w^z-1)$ \cite{ReviewHB,BookBH,BookSA}. 
We will devote the remaining part of this section to verify that $h(\ell,t)$ and $h(r,t)$ behave as conjectured in Eq.\ (\ref{hzxi})  for the two-dimensional percolation aggregate.

\subsection{Mortality function for the two-dimensional percolation aggregate: Chemical distance}
\label{subsec:MortalityChemical}

We start by estimating the mortality
function $h(\ell,t)$ when the ``circular'' absorbing boundary is placed at the chemical
distance $\ell$ from the starting site.
The numerical evaluation of this quantity is performed
by the Chapman-Kolmogorov method (also called the exact enumeration method
\cite{ReviewHB,BookBH}).
The circular boundary of absorbing traps is simulated
by a set of special sites belonging to the cluster that absorbs all the probability
density that enters them without giving back any probability to their neighbors.
In the simulation we locate these boundaries at distances $\ell=40, 100, 160$, 
and perform three experiments of absorption for $t=0,1,\dots,t_{\text{max}}$ with $t_{\text{max}}=1000$
on every aggregate built, one for each boundary.
The resulting mortality function
is averaged over $2000$ realizations of the percolation aggregate, which are generated by the Leath method \cite{BookBH,Leath}

\begin{figure}
\includegraphics{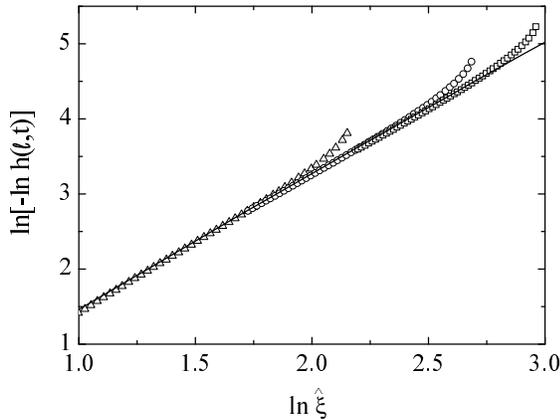}
\caption{\label{mortQui}
Plot of $\ln [-\ln h(\ell,t) ]$ versus $\ln \hat{\xi}$
for the  two-dimensional incipient percolation
cluster. The radii of the absorbing ``circular'' boundaries are
 $\ell=40$ (triangles), $\ell=100$ (circles) and $\ell=160$ (squares). 
 The line represents the function
$h(\ell,t)= \hat{A} \hat{\xi}^{-\protect{\hat{\mu}} \protect{\hat{v}}}
     \exp (-\hat{c}\protect{\hat{\xi}}^\protect{\hat{v}})$
with $\hat{A}=1$, $\hat{\mu}=-0.4$, $\hat{c}=0.9$ and $\hat{v}=d_w^{\ell}/(d_w^{\ell}-1)=1.714$.
}
\end{figure}

In Fig.\ \ref{mortQui} we plot $\ln[-\ln h(\ell,t)]$ versus
$\ln \hat{\xi}$ with $\hat \xi \equiv \ell/t^{1/d_w^\ell}$ and $d_w^{\ell}=2.4$.
[Hereafter we will put the symbol $\wedge$  ( $\sim$\,)
over quantities corresponding to the chemical  (Euclidean) distance.]
The value of $d_w^{\ell}$  was taken from one of our previous
 works \cite{SNTFrac} and  is in agreement with the values reported
by other authors \cite{ReviewHB,BHR,Majid}.
If the conjecture in
Eq.\ (\ref{hzxi}) is right we should observe  the linear behavior $\ln[-\ln
h(\ell,t)] \sim \ln \hat{c}+\hat{v} \ln \hat{\xi}$.
Certainly the 
plots seem linear  in Fig.\ \ref{mortQui} except for a portion in the range of large 
$\hat{\xi}$  where the curves deviate upwards. 
This is a finite size effect (already analyzed in the case of the
two-dimensional Sierpinski gasket in Ref. \cite{Propag})
associated with the existence of a minimum arrival time
corresponding to a random walker who travels ``ballistically''
along a chemical path from the origin to the absorbing boundary,
which in turn implies a maximum available value of $\hat{\xi}$ in
the simulations: $\hat{\xi}_{\text{max}} =\ell^{1-1/d_w^\ell}$.
This value is around $\hat{\xi}_{\text{max}} \simeq 2.15$
for $\ell=40$ and $\hat{\xi}_{\text{max}} \simeq 2.96$ for $\ell=160$.
This apparently means that the reliable interval for numerical
fitting is larger for $\ell=160$ but we must also take into
account that the minimum value of $\xi$ attained in the
simulations is $\ell/t_{\text{max}}^{1/d_w^{\ell}}$ which is
proportional to $\ell$ (in our simulations $t_{\text{max}}=1000$).
For this reason, in order to carry out the numerical fit, we have concatenated the simulation data for which the plots are almost  linear.
The result $h(\ell,t) = \hat{A}
\hat{\xi}^{-\hat{\mu}\hat{v}} \exp( -\hat{c} \hat{\xi}^{\hat{v}})$ with $\hat{A}=1$, $\hat{c}=0.9$, $\hat{v}=d_w^\ell/(d_w^\ell-1) \simeq 1.714$ and $\hat{\mu}=-0.4$  is also plotted in  Fig.\ \ref{mortQui}. The agreement is excellent.

 \subsection{Mortality function for the two-dimensional percolation aggregate:  Euclidean distance}
 \label{subsec:MortalityEuclidean}

\begin{figure}
\includegraphics{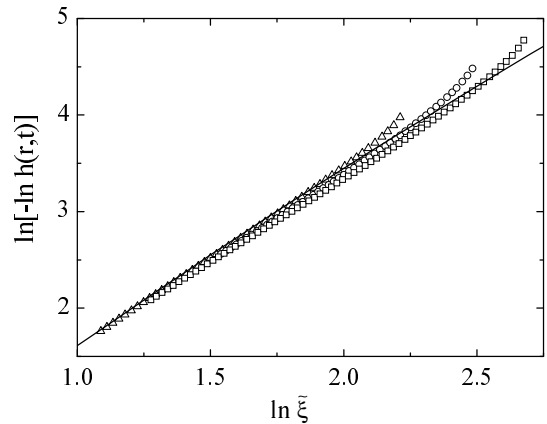}
\caption{\label{mortEuc}
Plot of $\ln [-\ln h(r,t) ]$ versus $\ln \tilde{\xi}$ for  $r=35$ (triangles),
$r=55$ (circles), $r=75$ (squares). The line represents the function
$h(r,t)= \tilde{A} \tilde{\xi}^{-\protect{\tilde{\mu}} \protect{\tilde{v}}} \exp (-\tilde{c} \tilde{\xi}^\protect{\tilde{v}})$ with 
$\tilde{A}=1$, $\tilde{\mu}=-1.8$, $\tilde{c}=1.65$ and $\tilde{v}=d_w/(d_w-1)=1.56$.
}
\end{figure}

We also simulated the mortality function $h(r,t)$ in the percolation
aggregate when the circular absorbing boundary has a Euclidean radius $r$. 
The analysis of these results parallels those of the previous subsection. In Fig.\  \ref{mortEuc}  we have plotted
the simulation results for the double logarithm of the
mortality function $\ln[-\ln h(r,t)]$ versus $\ln \tilde{\xi}$ (where
$\tilde{\xi}=r/t^{1/d^w}$) for circular boundaries of radii $r=35, 55, 75$, with $t_{\text{max}}=1000$, and $2000$ aggregates to perform the average. 
The anomalous diffusion coefficient for that time
range was taken as $d_w=2.8$ in agreement with that obtained by us in a previous work  \cite{SNTFrac}. 
In the long time limit, a slightly greater value
$d_w=2.87$ has been found \cite{ReviewHB}, but as our simulations are restricted to $t \le 1000$ it is more reasonable to consider the (effective) value $d_w=2.8$ which better represents the diffusive behavior in this time range.

\begin{figure}
\includegraphics{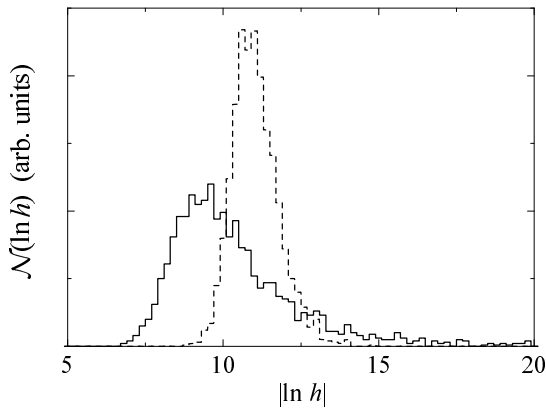}
\caption{\label{histograma}
Distribution of the mortality function with a circular trapping boundary,  $h(z,t)$, over 2000 realizations of
the  two-dimensional incipient percolation aggregate. We plot the histogram
${\cal N}[\ln h(z,t)]$ versus $\vert \ln h(z,t) \vert$  for  $t=1000$,   $z=r=45$ (solid line) and  $z=\ell=80$  (dashed line).
The distribution  is clearly wider for the Euclidean case than for the chemical case,
although to a quite noticeably lesser extent than for the site mortality function \protect\cite{SNTFrac} and
the propagator \protect\cite{BookBH,BHR}.
}
\end{figure}

In contrast with the results for the chemical boundary case, now we get a poorer collapse to a single line for the curves corresponding to different values of $r$.
This we attribute to the fact that a given Euclidean distance $r$ may correspond to
many chemical distances $\ell \ge r$ depending on the site and the particular
aggregate we are using in the simulation.
The result is that,  in contrast with their chemical counterparts,
the propagator, $P(r,t)$ \cite{BHR}, the mortality function for
a single trap site \cite{SNTFrac}, and the mortality function  for
 an absorbing spherical boundary, $h(r,t)$  (see Fig.\ \ref{histograma}), exhibit a  broad distribution over the
different  percolation aggregate realizations.
As a consequence, statistical precision requires more substrate averaging in the
case of quantities defined in terms of Euclidean distance than quantities
referred to the more natural chemical distance.
Anyway, proceeding as in Sec.\ \ref{subsec:MortalityEuclidean},  we linked
the simulation results for $r=35, 55, 75$ where the plot is almost
linear in order to perform an overall fit.
Assuming the functional form of Eq.\ (\ref{hzxi})  we find  that the data can be roughly described by 
$h(r,t) = \tilde{A}
\tilde{\xi}^{-\tilde{\mu}\tilde{v}} \exp( -\tilde{c} \tilde{\xi}^{\tilde{v}})$ with
$\tilde{A}=1$, $\tilde{\mu}
= -1.8$, $\tilde{c} =1.65$ and $\tilde{v} \equiv d_w/(d_w-1) \simeq 1.56$ (see Fig. \ref{mortEuc}).

 \section{ORDER STATISTICS ON A STOCHASTIC FRACTAL LATTICE}
\label{sec:FormulasAsintoticas}

 Now, with the  functional form for $h(z,t)$ found in the preceding section, we can proceed with the evaluation of  $\langle t^m_{j,N} \rangle$ by means of the asymptotic
technique already developed for Euclidean and deterministic fractal substrates. Only the key steps in the calculation
will be outlined. The interested reader is referred to previous
references for the details \cite{WSL,PRLYuste,YAL}.

We will find  it convenient in this section to
write the mortality function as follows:
\begin{equation}
\label{hzt}
h(t)\equiv h(z,t) \sim a(z) t^{\mu \beta} \exp \left[ - (t_0/t)^{\beta} \right]
\left\{ 1+\bar h_1 t^{\beta} \ldots \right\}
\end{equation}
where $a(z)=A z^{-\mu v}$,    $\beta=v/d_w^z$,
$t_0(z)=c^{1/\beta} z^{d^z_w}$ y $\bar h_1=h_1 z^{-v}$.
The generating function of the $m$th moment of the $j$th passage time, ${\cal U}_{N,m}(\zeta)=
\sum_{j=1}^N \langle t_{j,N}^m \rangle \zeta^{j-1}$, can be written as \cite{WSL}
\begin{equation}
\label{calUzeta}
{\cal U}_{N,m}=\frac{m}{1-\zeta} \int_0^\infty \, dt \, t^{m-1} \left\{  [1- h(t) + h(t) \zeta]^N-
\zeta^N \right\} .
\end{equation}
Dropping the $\zeta^N$ term we get
\begin{equation}
\label{Uzeta}
U_{N,m} \equiv \frac{m}{1-\zeta} \int_0^\infty \, dt \, t^{m-1} \exp \left\{ N \ln \left[
1-h(t) (1-\zeta) \right] \right\} .
\end{equation}
The point is that ${\cal U}_{N,m}(\zeta)$ and $U_{N,m}(\zeta)$ have the same Taylor series
expansion up to the term of order $\zeta^{N-1}$, so that $\langle t_{j,N}^m \rangle$ can
also be estimated through the evaluation of this pseudo-generating function
$U_{N,m}(\zeta)$.
 Proceeding as in Refs.\ \cite{WSL,PRLYuste,YAL}, one gets:
 \begin{widetext}
\begin{eqnarray}
\label{UNmexp}
U_{N,m}(\zeta) &=& \frac{t_0^m}{1-\zeta} \frac{1}{\ln^\alpha
\lambda} \left\{1+\frac{\alpha \left(\mu \ln \ln
\lambda-\gamma\right)}{\ln \lambda}+ \frac{\alpha}{2 \ln^2 \lambda}
\left[(1+\alpha)\left( \frac{\pi^2}{6}
+\gamma^2 \right)+2 \mu \gamma-2 \bar h_1 t_0^\beta \right. \right. \nonumber \\
\noalign{\smallskip}
& &\left. \left. -2 \mu \left( \mu+(1+\alpha) \gamma \right) \ln \ln \lambda
+(1+\alpha) \mu^2 \ln^2 \ln \lambda\right]+\mathit{O} \left(
\frac{\ln^3 \ln \lambda}{\ln^3 \lambda} \right) \right\}
\end{eqnarray}
\end{widetext}
where $\alpha\equiv m/\beta$ and $\gamma \simeq 0.577215$
is the Euler constant.

Once the generating function is known, the escape times and their moments are
calculated straightforwardly because $\langle t_{j,N}^m \rangle$ is simply
the coefficient of $z^{j-1}$ in the Taylor series expansion of
$U_{N,m}(\zeta)$. Therefore, the $m$th moment of the first passage time of
the first out of $N \gg 1$ diffusing particles is equal to $U_{N,m}(0)$, i.e.,
\begin{widetext}
\begin{eqnarray}
\label{t1Nm}
\langle t_{1,N}^m \rangle &\sim& \frac{t_0^m}{\ln^\alpha
(\lambda_0 N)} \left\{1+\frac{\alpha \left(\mu \ln \ln
\lambda_0 N-\gamma\right)}{\ln \lambda_0 N}+ \frac{\alpha}{2 \ln^2 (\lambda_0 N)}
\left[(1+\alpha)\left( \frac{\pi^2}{6}
+\gamma^2 \right)+2 \mu \gamma-2 \bar h_1 t_0^\beta \right. \right. \nonumber \\
\noalign{\smallskip}
& &\left. \left. -2 \mu \left( \mu+(1+\alpha) \gamma \right) \ln \ln \lambda_0 N
+(1+\alpha) \mu^2 \ln^2 \ln \lambda_0 N \right]+\mathit{O} \left(
\frac{\ln^3 \ln \lambda_0 N}{\ln^3 \lambda_0 N} \right) \right\}
\end{eqnarray}
\end{widetext}
with $\lambda_0=a t_0^{\mu \beta}=Ac^{\mu}$ and $\alpha=m/\beta=mv/d^z_w$ as before.
Notice that the main term of this expression for $m=1$  agrees with the result given in \cite{PREDK} if we use  $v=d_w^z/(d_w^z-1)$.

The calculation
of $\langle t_{j,N}^m \rangle$ for $j > 1$ is more involved. We begin with
the identity $\ln^n (1-z)=n! \sum_{i=n}^\infty (-1)^i S_i(n) z^i/i!$, where
$S_i(n)$ are the Stirling numbers of the first kind \cite{Abra}.
Using this relation, the expansions of $1/\ln^\alpha \lambda$ and similar terms in
Eq.\ (\ref{UNmexp}) in powers of $1/\ln \lambda_0 N$ can be found \cite{PRLYuste}.
After some algebra we have
\begin{equation}
\label{tjNm}
\langle t_{j,N}^m \rangle \sim \langle t_{1,N}^m \rangle+
\frac{t_0^m \alpha}{\ln^{\alpha+1} (\lambda_0 N)} \sum_{n=1}^{j-1}
\frac{\Delta_n (\alpha)}{n}
\end{equation}
where $j=2,3,\ldots$ and
\begin{eqnarray}
\Delta_n(\alpha)&=&1+\frac{\alpha+1}{\ln \lambda_0 N} \left[
(-1)^n \frac{S_n(2)}{(n-1) !}+\mu \ln \ln \lambda_0 N \right. \nonumber \\
   &&\left. -\frac{\mu}{\alpha+1}
-\gamma \right]+{\cal O}\left( \frac{\ln^2 \ln \lambda_0 N}{\ln^2 \lambda_0 N}
\right) .
\label{Dnalpha}
\end{eqnarray}

Finally, we will quote the main term of the asymptotic expression for
the variance, $\sigma_{j,N}^2=\langle t_{j,N}^2 \rangle-\langle t_{j,N} \rangle^2$,
which is easily derived from Eqs. (\ref{t1Nm}) and (\ref{tjNm}) yielding
\begin{equation}
\label{sigmajN2}
\sigma_{j,N}^2=\frac{t_0^2}{\beta^2}
\displaystyle\frac{d_j}{(\ln \lambda_0 N)^{2+2/\beta}}
\left[ 1+ \mathit{O}\left( \frac{\ln^3 \ln \lambda_0 N}{\ln \lambda_0 N}
\right)   \right]    \;    
\end{equation}
with
\begin{equation}
d_j=\left[
\frac{\pi^2}{6}-\left( \sum_{n=1}^{j-1} \, \frac{1}{n} \right)^2
+2 \sum_{n=1}^{j-1} \, (-1)^n \frac{S_n(2)}{n !} \right]
\end{equation}
and  $j=1,2,\ldots$.  We use the convention that the sums are equal
to zero when the upper limit is zero. It is clear that the main and first
corrective terms of $\langle t_{j,N}^2 \rangle$ are equal to those of $\langle
t_{j,N} \rangle^2$, so that only the difference between their second corrective
terms contributes to the {\em main} term of the variance. 
For the sake of comparison with the simulation results it is more convenient to consider the quotient $\langle t_{j,N}
\rangle / \sigma_{j,N}$ whose expression is given by
\begin{equation}
\label{tjNdsjN}
\frac{\langle t_{j,N} \rangle}{\sigma_{j,N}} =
\beta d_j
 \ln N
\left[ 1+\mathit{O} \left( \frac{\ln^3 \ln N}{\ln N} \right) \right],
\end{equation}
where $d_1=(\pi^2/6)^{-1/2}$, $d_2=(\pi^2/6-1)^{-1/2}$, $d_3=
(\pi^2/6-5/4)^{-1/2}$,\ldots.

Let us now compare the theoretical predictions in Eqs.\
(\ref{t1Nm}), (\ref{tjNm}) and (\ref{tjNdsjN}) with simulation results in
the two-dimensional percolation aggregate when an  absorbing ``circular''  boundary  is placed either at  a given  chemical distance  $z=\ell$ or at a given Euclidean distance $z=r$.
 
\subsection{Absorbing boundary at a given chemical distance}

\begin{figure}
\includegraphics{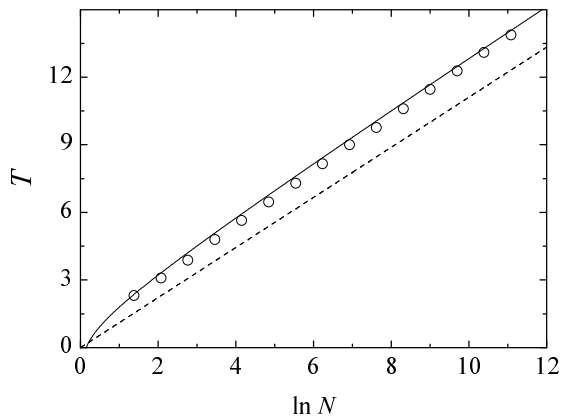}
\caption{\label{t1NQui}
Plot of $T\equiv \left[ \langle t_{1,N} \rangle/\ell^{d_w^\ell} \right]^{-\delta}$
 versus $\ln N$ ($N=2^2$, $2^3$, \ldots, $2^{16}$) for the two-dimensional incipient percolation
aggregate where $\delta=1/(d_w^\ell-1)$. The circles are  simulation
results for $\ell=50$ and the lines are  the zeroth-order (broken line) and first-order (solid line) asymptotic approximations with $\hat{c}=0.9$, $\hat{v}=1.714$ and $\hat{\mu}=-0.4$.}
\end{figure}

 The first passage time of the first few random walkers out of a set of $N=2^i$, $i=2,3,\ldots,16$
to a ``circular''  boundary of chemical radius $\ell=50$ was simulated on  $2000$ aggregates. 
In Fig.\ \ref{t1NQui} we plot the scaled simulation results of 
 $[ \langle   t_{1,N} \rangle/\ell^{d_w^{\ell}} ]^{-\delta}$, with $\delta=1/(d_w^{\ell}-1) $, versus $\ln N$ for
$\ell=50$ and compare them with the theoretical predictions of Eq.\ (\ref{t1Nm}), namely,
\begin{eqnarray}
\label{first} 
\left[ \frac{\langle t_{1,N}
\rangle}{\ell^{d_w^{\ell}}} \right]^{-\delta} &\approx&
\left[ \frac{\ln N}{\hat{c}}
\right]^{\frac{\delta d_w^\ell }{\hat{v}}} \left\{1 \right.\nonumber \\
&&
\left.-\frac{\delta d_w^\ell}{\hat{v}}\, \frac{\hat{\mu}
\ln \ln N -\gamma-\ln \hat{A} \hat{c}^{\hat{\mu}}}{\ln N} \right\}
\end{eqnarray}
where  $\delta=1/(d_w^\ell-1)$. 
Notice that $\left( \langle t_{1,N} \rangle/\ell^{d_w^{\ell}} \right)^{-\delta}$ depends linearly on $\ln N$ if $\hat{v}=d_w^\ell/(d_w^\ell-1)$. Figure \ref{t1NQui} shows that this is indeed the case.
 
\begin{figure}
\includegraphics{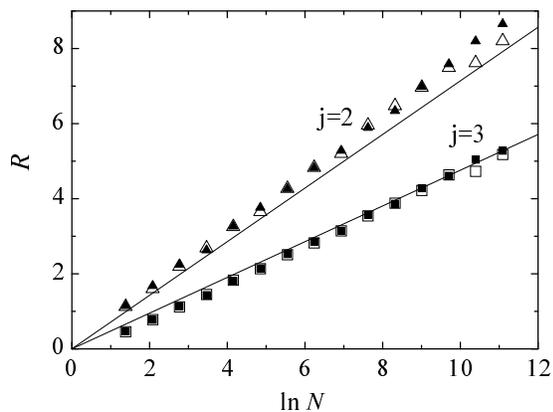}
\caption{\label{tjNQui}
Plot of $R\equiv \langle t_{1,N} \rangle/\langle t_{j,N}-t_{1,N} \rangle$ versus
$\ln N$ for the two-dimensional incipient percolation aggregate with $j=2$ and
$j=3$. The hollow [filled] symbols are the simulation results for $\ell=50$ [$\ell=80$]
which superpose closely except for large $N$. 
The lines correspond to the zeroth-order theoretical
prediction $\ln(N)/[(d_w^\ell-1)\sum_{n=1}^{j-1}(1/n)]$  with $d_w^\ell=2.4$.
}
\end{figure}

From Eq.\ \eqref{tjNm} one gets 
\begin{equation}
\label{tjNplot}
R\equiv \frac{\langle t_{1,N} \rangle}{\langle t_{j,N}-t_{1,N}\rangle} \sim
\left( \sum_{n=1}^{j-1} \frac{1}{n}\right)^{-1} \frac{\hat{v}}{d_w^\ell} \ln N.
\end{equation}
so that $R$ should be linear in $\ln N$ and independent of the radius of the chemical boundary $\ell$.  This is confirmed in Fig.\  \ref{tjNQui} where this quotient is plotted versus $\ln N$ for $j=2, 3$ and $\ell=50, 80$. 
A good superposition of the simulation data for $\ell=50$ and $\ell=80$ is observed. 
The simulation results are in good agreement with the prediction of Eq.\ (\ref{tjNplot}) with  $\hat{v}=d_w^\ell/(d_w^\ell-1)$ and $d_w^\ell=2.4$.

\begin{figure}
\includegraphics{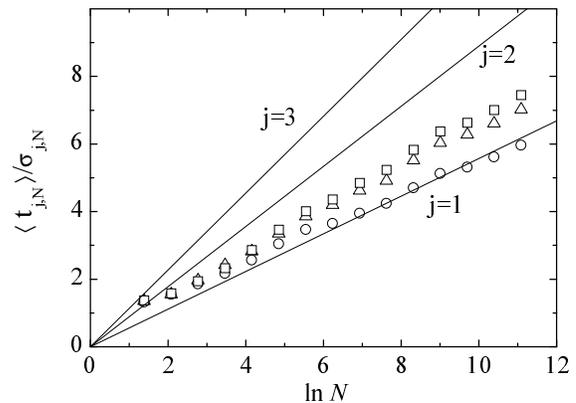}
\caption{\label{varjN}
Plot of $\langle t_{j,N} \rangle/\sigma_{j,N}$ versus $\ln N$ for the
two-dimensional incipient percolation aggregate  with $\ell=50$. The circles, triangles and
squares denote the simulation results for $j=1$, $2$ and $3$, respectively. The
 lines are the theoretical predictions with $d_w^\ell=2.4$ and
$\hat{v}=d_w^\ell/(d_w^\ell-1) = 1.714$.}
\end{figure}

\begin{figure}
\includegraphics{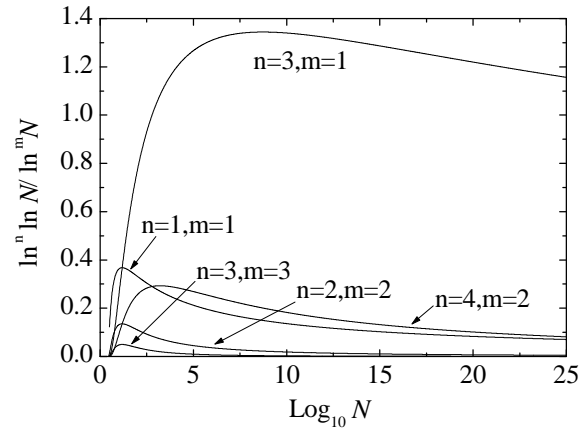}
\caption{\label{lnnNlnmN}
Plot of $\ln^n \ln N/\ln^m N$ versus the decimal logarithm $\text{Log}_{10} N$
for several pairs $(n,m)$. These functions gauge  the relative error of different finite-order asymptotic approximations to the order-statistics quantities.}
\end{figure}

Finally, in Fig.\  \ref{varjN} we show the quotient between the
average lifetime of the $j$th particle and its variance, $\langle t_{j,N}
\rangle/\sigma_{j,N}$, versus $\ln N$ for the first, second and third random walkers to arrive at the chemical boundary with $\ell=50$.
This is a remarkable figure that shows the crucial importance of the corrective terms on the order statistics formulae.
Except for the first walker ($j=1$) we find striking discrepancies between the zeroth-order (main term) of the
asymptotic expression  in Eq.\ (\ref{tjNdsjN}) and the simulation results.
We attribute these discrepancies to the 
$\mathit{O} \left[ (\ln^3 \ln N)/\ln N \right]$ corrective terms not considered in the zeroth-order asymptotic expression. Let us explain why.
In  Fig.\ \ref{lnnNlnmN} we plot the function $\ln^n \ln N/\ln^m N$  for several values of $n$ and $m$.  
One sees that  $\ln^3 \ln N/\ln N$ cannot be neglected versus 1 even for values of $N$  larger than the Avogadro's number. 
Therefore, it is not strange to find a poor agreement between the zeroth-order asymptotic expression and the simulation results.  
(In fact, what is truly unexpected is the good agreement of the zeroth-order theoretical prediction  with simulation  results for $j=1$.) 
Notice  that if we had calculated another corrective term in the asymptotic
expansion  (\ref{UNmexp}), then the term neglected in Eq.\ (\ref{tjNdsjN}) would be of
order $\ln^4 \ln N/\ln^2 N$, which, as shown in Fig.\ \ref{lnnNlnmN},  would likely be much
smaller than the $\ln^3 \ln N/\ln N$ corrections.
Unfortunately, corrective terms of order $\ln^3 \ln N/\ln N$ involve parameters such as $h_1$ [see Eq.\ (\ref{hzxi})] which are very difficult to estimate relying only on simulation results.  
One also observes in Fig.\ \ref{lnnNlnmN} that the relative errors of the zeroth- and first-order approximations to $\langle t_{j,N} \rangle$ committed by ignoring terms of the form $\ln \ln N/\ln N$ and $\ln^2 \ln N/\ln^2 N$, respectively, will likely be small.
This explains the good agreement between theory and simulation for lifetimes shown in Figs.\ \ref{t1NQui} and \ref{tjNQui}.

\subsection{Absorbing boundary at a given Euclidean distance}

\begin{figure}
\includegraphics{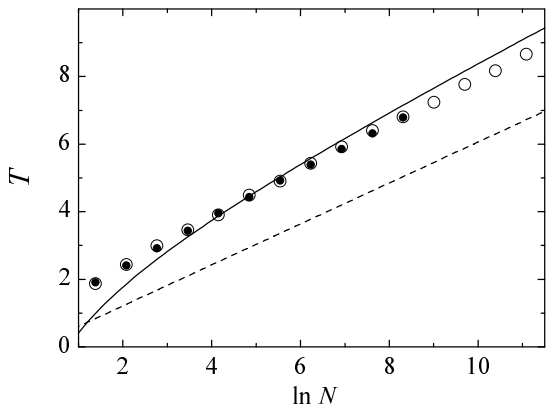}
\caption{\label{t1NEuc}
Plot of $T=\left[ \langle t_{1,N} \rangle/r^{d_w} \right]^{-\delta}$
 versus $\ln N$ for the two-dimensional incipient percolation
aggregate, where  $N=2^2$, $2^3$, \ldots, $2^{16}$ and $\delta=1/(d_w-1)$. The filled [hollow] circles are the simulation
results for $r=50$ [$r=75]$.
The broken and solid lines are the zeroth-order and first-order asymptotic approximations, respectively, 
 with $\tilde{c}=1.65$, $\tilde{v}=1.56$ and $\tilde{\mu}=-1.85$.}
\end{figure}

\begin{figure}
\includegraphics{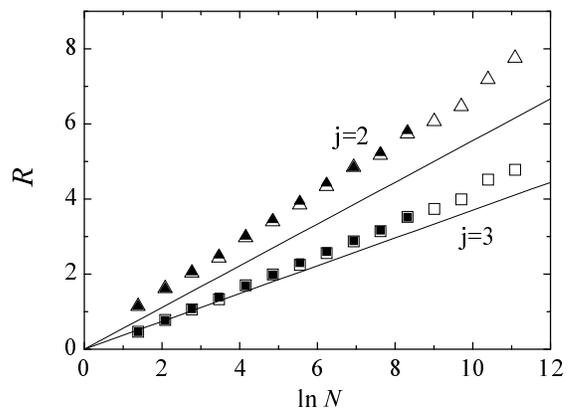}
\caption{\label{tjNEuc}
The same as Fig.\ \protect\ref{tjNQui} but for circular boundaries 
with Euclidean radius $r=50$ (filled symbols) and $r=75$ (hollow symbols). 
The lines correspond to the zeroth-order theoretical
prediction $\ln(N)/[(d_w-1)\sum_{n=1}^{j-1}(1/n)]$  with $d_w=2.8$.
}
\end{figure}

We have also simulated the order-statistics of the arrival of $N$ random walkers at a circular boundary of Euclidean radius $r$. 
The analysis of these results parallels  those of the previous subsection.
The numerical results for $[ t_{1,N}/r^{d_w} ]^{-1/(d_w-1)}$ and
$\langle t_{1,N} \rangle/\langle t_{j,N}-t_{1,N} \rangle$ versus $\ln N$ are
plotted in Figs.\ \ref{t1NEuc} and \ref{tjNEuc}, respectively.
The radii of the circular boundary are $r=50$ and $r=75$, and we used $2000$ aggregates
to perform the substrate average.
The theoretical predictions of Eqs.\ (\ref{first}) and (\ref{tjNplot})
 [replacing $\ell$ by $r$,  $d_w^\ell$ by $d_w$, $\delta$ by $1/(d_w-1)$ and 
$\wedge$ by $\sim$]
lead to an agreement with numerical results that is not as good as that obtained when the chemical distance was used (see Figs.\ \ref{t1NQui} and \ref{tjNQui}), although the linear behavior of $T$ and $R$ versus $\ln N$ and the superposition  of the simulation data for different values of $r$ is still there.
This slightly worse prediction is in accordance with the fact, discussed in Sec.\ \ref{subsec:MortalityEuclidean}, that the collapse of the numerical mortality function for different distances $r$ to a single curve is not so nearly perfect as when chemical distances are used.

\section{CONCLUSIONS AND REMARKS}
\label{sec:Conclusions}
In this paper we studied the order statistics of the survival times (lifetimes or exit times)  of $N\gg 1$ independent random walkers that are put initially on a stochastic lattice at the center of a ``spherical'' absorbing boundary of radius $z$.
We found the moments  $\langle t_{j,N}^m \rangle$ of the lifetime of the $j$th
random walker of the set of $N$  in terms of an asymptotic series that decays mildly in powers of $1/\ln N$.
These theoretical results were compared with numerical simulations for the two-dimensional incipient percolation aggregate with the absorbing boundary  placed either  at a given chemical distance $z=\ell$ or at a given Euclidean distance $z=r$. 
The agreement found was reasonable and in accordance with the large size of the asymptotic corrective terms. 

 The theoretical approach relies on the knowledge of the mortality function  $h(z,t)$
for short times.  
We assumed that the functional form of the short-time mortality function for a stochastic lattice averaged over all realizations coincides with that of the mortality function for Euclidean lattices and  for the class of deterministic  fractal lattices whose sites are isolated by their nearest neighbors.
This is the main assumption of this paper and we  showed that  it is compatible
with our simulation results for the two-dimensional percolation aggregate with  Euclidean and chemical absorbing circular boundaries. This allows us to describe the order statistics of the diffusion process on many kinds of substrates (namely, Euclidean media, and deterministic and {\em stochactic} fractals) with the same approach and formulae.

A useful feature of the order statistics description is that it allows one to infer properties of the diffusive system (diffusion constant, number of diffusing particles, concentration of traps,  effective dimension of
the diffusive substrate,\ldots) from  the analysis of the behavior of those
particles that are trapped first.   
This could be an advantage when it is impractical or impossible to wait until  the entire reaction is over. 

Finally, it should be noted that our simulation results for the stochastic fractal we have studied (the incipient percolation aggregate) are insufficient to perform a completely reliable numerical fit to the general form of the mortality function $h(z,t)$  with its dominant exponential {\em and} subdominant power-law terms simultaneously. 
Indeed, even to find  the true exponent value $v$ of the {\em dominant} term for the propagator is not at all easy as past controversies about that true value reveal.
Moreover,  the value of a hypothetical power-law subdominant term
[equivalent to the term $\xi ^{-\mu v}$ of Eq.\ (\ref{hzxi})] is still a matter of discussion
 \cite{JPAYuste,Propag,propuestaspropagador}.
Hence, one must expect that to find a completely reliable description of the
short-time mortality function will be similarly difficult, and this task will therefore require more detailed simulations on larger lattices, with longer simulation times and, of course, averaging over many more realizations of the lattice than those we have used here.
This is simply too much for a common workstation, but we think the effort on more powerful systems  would be worthwhile:  first, because of the interest in the percolation aggregate as a model of disordered media \cite{ReviewHB,BookBH}, and,
second, because  simulation of the mortality function (with circular boundary) is easier
than that of the propagator and, since the asymptotic coefficients  and exponents of these two functions are related, 
estimation of these parameters for the mortality function would help to solve the
long-standing controversy on the exact asymptotic form of the propagator in fractal media.

\acknowledgments
This work has been supported by the Miniterio de Ciencia y Tecnolog\'{\i}a (Spain) through Grant No. BFM2001-0718.
SBY is also grateful to the DGES (Spain) for a sabbatical grant (No.\ PR2000-0116) and to
Prof.\ K.\ Lindenberg and the Department of Chemistry of the University of California San Diego for their  hospitality.



\begin{thebibliography}{99}
\bibitem{surviEucDetFrac} For Euclidean media, the survival probability  can be found by solving the diffusion equation  with an absorbing spherical boundary placed at the Euclidean distance $r$.
For the class of deterministic fractal media discussed in \cite{PRLYuste},
 the survival probability is found by renormalization techniques.
\bibitem{conjecture}A key fact that supported the conjecture was  that  the asymptotic relation for  $S_N(t)$  obtained in Ref.\ \protect{\cite{PRAHL}} for disordered fractals can also be re-obtained from Eq. (\protect{\ref{eq:soluformal}}).  However, it was found later \protect{\cite{SNTFrac,PREDK}}  that that relation for  $S_N(t)$  is not valid for disordered fractals.
\end{thebibliography}
\end{document}